\documentstyle[prcrc,11pt,fleqn]{article}

\title{A fluid mechanical explanation of dark matter}

\author{Carl H. Gibson\address{Departments of Applied Mechanics
and Engineering Sciences and Scripps Institution of Oceanography, 
        University of California at San Diego \\ 
        MC 0411, La Jolla, CA 92093-0411, USA}%
        \thanks{cgibson@ucsd.edu; http://www-acs.ucsd.edu/ $\tilde{}$  ir118}}

\begin{document}
\maketitle

\begin{abstract}
Matter in the universe has become ``dark'' or ``missing'' through
misconceptions about the fluid mechanics of 
gravitational structure formation.  The linear Jeans (1902, 1929)
theory gives vast overestimates
(by factors of trillions or more) for the initial condensation
mass of baryonic matter, and vast underestimates (by factors of
trillions or more) for the initial condensation mass of the
non-baryonic matter.  Gravitational condensation occurs on non-acoustic
density nuclei at the largest Schwarz length
scale $L_{ST},L_{SV},L_{SM},L_{SD}$
permitted by turbulence, viscous, or magnetic forces,
or by the fluid diffusivity.  Non-baryonic fluids have diffusivities
larger (by factors of trillions or more) than baryonic (ordinary)
fluids, and cannot condense to nucleate baryonic galaxy formation
as is usually assumed.  Baryonic fluids begin to condense in the plasma
epoch at about 13,000 years after the big bang to form proto-superclusters,
and form proto-galaxies by 300,000 years when the cooling plasma
becomes neutral gas.  Condensation occurs at small planetary masses to form
``primordial fog particles'' from nearly all of the primordial gas by the new
theory, Gibson (1996), supporting the Schild (1996) conclusion from
quasar Q0957+651A,B microlensing observations that the mass of the lens 
galaxy is dominated by ``rogue planets ... likely to be the missing mass''. 
Non-baryonic dark matter condenses on superclusters at scale $L_{SD}$
to form massive super-halos. 
\end{abstract}

\section{Introduction}

Gravitational structure formation is intrinsically non-linear.  
As matter is pulled toward a density maximum by gravity
to form a condensate and pulled away by gravity
from a density minimum 
to form a void, both gravitational forces are
increased by the resulting redistribution of mass
so that the mass flow rates are accelerated.  This is
a classical positive feedback process.
Linear theories are notoriously
misleading when applied to non-linear processes.  
Fluid mechanics provides a familiar example, where 
steady laminar solutions
for the momentum equations with non-linear terms
neglected are vastly different from the random
turbulent motions actually observed when
the Reynolds number (the ratio
of inertial to viscous forces) is large.

Jeans \cite{jns02,jns29} applied a linear perturbation
stability analysis to the problem of gravitational structure
formation in a motionless gas with uniform properties.  
Neglecting non-linear terms of the momentum and 
density conservation equations reduces the problem to one
of acoustics.  Sound wave-crests provide density maxima that can nucleate
gravitational condensation only if a wave-length $\lambda$
propagation time $\tau_{\lambda} = \lambda/V_S$  
is greater than the ``free fall'' gravitational
time scale $\tau_G = (\rho G)^{-1/2}$, where $V_S$ is the speed
of sound in the gas,
$\rho$ is the gas density
and $G$ is Newton's gravitational constant
$6.7\times10^{-11}$ $ \rm m^3 \> kg^{-1} \> s^{-2}$. 
Thus, acoustic density
perturbations with sizes $L \ge L_J = V_S/(\rho G)^{1/2}$ 
can grow, but those smaller cannot.
This Jeans criterion for gravitational instability 
has been universally adopted in astrophysics
and cosmology, with modifications to take into account
effects of general relativity and the rapid expansion
rate of the early universe when the first structures
formed \cite {pbl93,kol94}. 
However, the Jeans criterion is unreliable.  Its widespread application 
is responsible for the present dark
matter paradox. 

Jeans asserted that acoustic
density maxima with $\lambda \ge L_J$ remain acoustic 
with velocity $V_S$ after condensation
begins.  This is one of many misconceptions in \cite {jns29},
including the claim that cores of galaxies consist of hot gas and not stars,
and are sources of matter flowing from other Universes. 
Actually, acoustic density nuclei must rapidly 
stop moving with sound speed $V_S$ as they accumulate 
zero momentum from the motionless ambient fluid.
Gibson \cite {gib96} shows that non-acoustic density nuclei
are absolutely unstable to gravitational structure formation
in a motionless fluid.  Most density maxima and minima are 
produced by turbulent mixing
of density.  They move approximately with the fluid velocity 
and have a characteristic length scale
$L_B = (D/\gamma)^{1/2}$ termed the Batchelor scale, where
D is the molecular diffusivity of the density and $\gamma$
is the rate-of-strain of the turbulence \cite {gib68}.  
Gravitational condensation on non-acoustic
density maxima and void formation from non-acoustic density
minima are limited either by the dominant fluid force or by molecular
diffusivity to length scales larger than the largest ``Schwarz'' length
scale $L_{SX}$.  Schwarz scales are derived in the following Section
by comparing dispersive forces with the force of gravity, and the
diffusion velocity with the gravitational velocity.  
The subscript ``X'' may be $T,V,M,D$ referring to either turbulent, 
viscous, magnetic or diffusive limitations to gravitational condensation.

\section{Theory}
A density nucleus with scale $L \ll L_J$ and mass
excess $M'$
in a large body of motionless, homogeneous gas is gravitationally 
unstable and will form a condensate, just
as the same sized density nucleus with
mass deficit $-M'$ is gravitationally unstable
and will form a void.  Both cases contradict Jeans' theory.   
If the gas is not motionless, gravitational forces
$F_G \approx \rho ^2 G L^4$ on scale L must be larger than
turbulence forces $F_T \approx \varepsilon ^{2/3} L^{8/3}$, viscous
forces $F_V \approx \rho \nu \gamma L^2$, or magnetic
forces $F_M \approx (H^2 /\rho) L^2$, where $\varepsilon$ is the
viscous dissipation rate, $\nu$ is the kinematic viscosity, $H$ is
the magnetic field, and $H^2 /\rho$ is the magnetic pressure.  Equality
occurs at the turbulent Schwarz scale 
$L_{ST} = \varepsilon ^{1/2} / (\rho G)^{3/4}$,
the viscous Schwarz scale $L_{SV} = (\nu \gamma / \rho G)^{1/2}$,
and the magnetic Schwarz scale
$L_{SM} = [(H^2 /\rho)/\rho G]^{1/2} = 
[(\chi_H /\varepsilon ^{1/3} )/\rho ^2 G]^{2/3}$, where
$\chi_H$ is the diffusive dissipation rate
of the magnetic field variance. From turbulent mixing 
theory, density fluctuations
on scale $L$
diffuse with velocity $D/L$ \cite {gib68}. Equating this to gravitational
velocities $L/\tau_G$ gives 
the Schwarz diffusive scale $L_{SD} = (D^2/\rho G)^{1/4}$.
Condensation of non-baryonic fluids on galactic
scales is impossible because $L_{SD} \gg L_{galaxy}$ \cite {gib96}.  

\section{Application of the theory}
By the new theory \cite {gib96}, gravitational decelerations to form
proto-superclusters and proto-galaxies begin in the plasma
epoch at 13,000 years after the big bang,
limited by viscous forces at scale $L_{SV}$.  
The initial conditions of the primordial hydrogen and helium
gas are constrained by the COBE satellite observation that
$\delta T/T \approx 10^{-5}$.  Turbulence levels were thus extremely
weak, with maximum rates-of-strain only slightly larger than
that of the expanding universe $\gamma = 1/t = 10^{-13} s^{-1}$.
The viscosity $\nu$ and diffusivity $D$ of the mixture were
about $10^{13}$ $\rm m^2 s^{-1}$, giving the condensation
mass $M_{SV} = {L_{SX} {}_{max}}^3 \rho = 
{L_{SV}}^3 \rho = 10^{23-24} \rm kg$.  The entire universe
of primordial gas condensed to form small planetary
mass objects (Mercury to Mars) 
termed ``primordial fog particles'', or PFPs.

\section{Observations}
Schild \cite {sch96} concludes that
his lensing galaxy consists mostly of small planetary objects
``likely to be the missing mass'', based
on continuous month-period fluctuations between
quasar image Q0957+651A,B light curves with a 1.1 year time delay.
Three observatories confirm the same time delay and the same fluctuating
(microlensed) signals \cite {gs98}. MACHO and EROS collaborations
exclude Galactic planetary masses as
the halo dark matter assuming
homogeneous spatial distributions of the 
objects \cite {alc97}. However, large 
clumping corrections
to star microlensing mass exclusion diagrams are required 
for $10^{-7} M_{\odot}$ objects 
because such small objects develop extreme spatial intermittency 
during their 15 billion years of nonlinear
gravitational accretion.  Such corrections could 
resolve the apparent observational inconsistency between
quasar-microlensing \cite {sch96} and 
star-microlensing \cite {alc97} studies of small
planetary mass objects as galaxy dark matter.

\section{Conclusions}
The Jeans criterion for gravitational instability is unreliable, and 
is replaced by a new criterion that density fluctuations in a gas
will experience gravitational condensation and void formation on
length scales larger than the maximum
Schwarz scale; that is, $L \ge L_{SX} {}_{max} $.  The baryonic 
``missing mass'' or ``dark matter'' consists of
those PFPs that have not aggregated to form stars.  Because
non-baryonic dark matter has $L_{SD}$ scales larger than
galactic scales it
is a negligible part of the galactic dark matter, but 
dominates supercluster dark matter as super-halos \cite {gib96}.

\end{document}